\documentclass[aps,prb,twocolumn,showpacs,superscriptaddress,groupedaddress]{revtex4-1}  % for review and submission
\usepackage{graphicx}  % needed for figures
\usepackage{dcolumn}   % needed for some tables
\usepackage{bm}        % for math
\usepackage{amssymb}   % for math
\usepackage{amsmath}
\usepackage{physics}
\usepackage{bm,color}
\usepackage{amssymb,amsfonts,amsmath}

\begin{document}

\title{Quantum spin Hall density wave insulator of correlated fermions}
\author{Gaurav Kumar Gupta}
\author{Tanmoy Das}
\affiliation{Department of Physics, Indian Institute of Science, Bangalore, India - 560012}

\date{\today}

\begin{abstract}
We present the theory of a new type of topological quantum order which is driven by the spin-orbit density wave order parameter, and distinguished by $Z_2$ topological invariant. We show that when two oppositely polarized chiral bands [resulting from the Rashba-type spin-orbit coupling $\alpha_k$, $k$ is crystal momentum] are significantly nested by a special wavevector ${\bf Q}\sim(\pi,0)/(0,\pi)$, it induces a spatially modulated inversion of the chirality ($\alpha_{k+Q}=\alpha_k^*$) between different sublattices. The resulting quantum order parameters break translational symmetry, but preserve time-reversal symmetry. It is inherently associated with a $Z_2$-topological invariant along each density wave propagation direction. Hence it gives a weak topological insulator in two dimensions, with even number of spin-polarized boundary states. This phase is analogous to the quantum spin-Hall state, except here the time-reversal polarization is spatially modulated, and thus it is dubbed quantum spin-Hall density wave (QSHDW) state. This order parameter can be realized or engineered in quantum wires, or quasi-2D systems, by tuning the spin-orbit couping strength and chemical potential to achieve the special nesting condition.
\end{abstract}
\maketitle

A topological state of matter can arise when two bands with opposite chirality are inverted across the Fermi level at odd number of time-reversal (TR) invariant momenta (TRIM).\cite{TIreviewTD,TIreviewSCZ,TIreviewCK} One of the prerequisites is thus to obtain a momentum dependence of the spin state or chirality, which is often triggered by the spin-orbit coupling (SOC). The inversion of the chirality between the bulk conduction and valence bands across the insulating band gap at the TRIM is protected by the TR symmetry, leading to a $Z_2$ topological insulator (TI). At the boundary, both chiral states meet at the TRIM with gapless edge or surface states. Within the Dirac Hamiltonian notation, the inverted bulk band gap (denoted by $m<0$) at the TRIM provides the negative Dirac mass, while the associated gapless boundary states. 

While strong quantum fluctuations or disorder are often detrimental to the band topology, they can conversely drive the inversion of the chiral bands with non-trivial topological properties. These states are not always defined by a Landau order parameter, rather distinguished by a topological invariant of the correlated electronic bands. Examples of such states include topological Mott,\cite{TMI,TAI} Kondo,\cite{TKI} and Anderson\cite{TDI} insulators. Antiferromagnetic order parameter can give a distinct topological class which breaks time-reversal and translation symmetries, but preserves their combinations.\cite{TAFM} To date, TIs have been realized in various non-interacting systems including HgTe/CdTe,\cite{HgTeTh,HgTeExp} InAs/GaSb~\cite{InAsGaSb} quantum wells for two-dimensional (2D) TIs, and Bi-based chalcogenides for 3D TIs.\cite{BiSbTh,BiSbExp,BiSeTh,BiSeExp,BiTeExp} SmB$_6$,~\cite{TKI,SmB6} and YbB$_6$~\cite{YbB6} have been extensively studied both theoretically and experimentally as potential candidates for topological Kondo insulators. 

\begin{figure}[ht]
\includegraphics[scale=0.14]{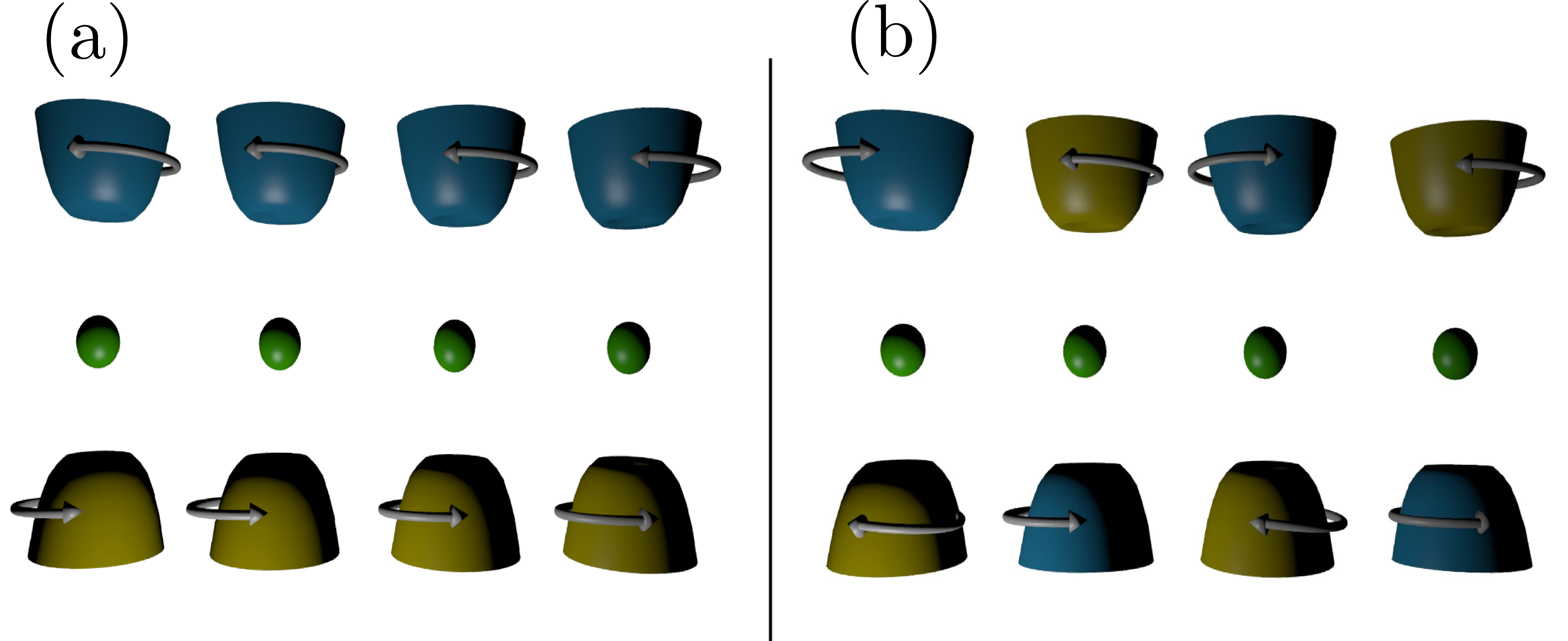}
\caption{(Color online) Distinction between a QSH and QSHDW insulator in real space. ({a}) A typical QSH insulator where all lattice sites have the same chirality in the valence band. ({b}) The QSHDW insulator where two sublattice sites have the opposite chirality in the valence band.}
\label{fig1}
\end{figure}

{\it Proposal:} We develop the theory of a Landau-type topological order parameter driven by staggered chiral band inversion. The order parameter arises from the translational symmetry breaking due to Fermi surface (FS) nesting between Rashba-type SOC (RSOC) split bands. Such nesting between opposite chiral states may occur in 2D systems or quantum wires of Bi, Pb, Sb, and similar elements in which both SOC and interaction are large.\cite{PbFS,PbNC} The nesting strength is enhanced with reduced system dimensionality and thickness.\cite{PbFS,PbNC} Our theory relies on a particular nesting vector ${\bf Q}\sim(\pi,0)$ or $(0,\pi)$, where the helicity of the RSOC $\alpha_{\bf k}=\alpha_R(\sin{k_y}-i\sin{k_x})$ (with $\alpha_R$ being the RSOC strength, and $k_x,~k_y$ are the crystal momenta) is reversed to $\alpha_{{\bf k}+{\bf Q}}=\alpha_{\bf k}^*$. This is the key feature responsible for modulated chiral band inversion. We find that as a Landau type order parameter develops due to this FS instability, it leads to a negative Dirac mass and insulating band gap. Along the direction of the nesting, we find that  correlated electronic bands are associated with non-trivial $Z_2 $ invariant, with spin-polarized zero-energy boundary states. Such a state can be compared with a non-interacting QSH insulator in 2D, with the distinction that here every alternative atoms possess opposite chirality in the same valence band, owing to translational symmetry breaking, as illustrated in Fig.~\ref{fig1}(b). Thus we call it a quantum spin-Hall density wave (QSHDW) insulator.

{\it Theory of QSHDW:} 
To develop the theory of QSHDW, we use a single band tight-binding model in a 2D lattice with RSOC. The FS nesting is generally known to increase as the dimensionality is reduced. For this reason, we use anisotropic tight-binding hoppings along the $x$- and $y$-directions ($t_x$ and $t_y$), so that the nesting at the wavevector $Q=(\pi,0)$ or $(0,\pi)$ can be monitored by changing the ratio $t_x/t_y$. The concept and formalism of the QSHDW is general for any dimension as long as the corresponding nesting wavevector allows for the chirality inversion at all given dimensions. We use a tight-binding dispersion with nearest neighbor hopping as $\xi_\textbf{k}=-2 \left[t_x\textrm{cos}(k_xa) +t_y \textrm{cos}(k_yb)\right] -\xi_{\rm F}$, where $\xi_{\rm F}$ is the chemical potential, and $a$ and $b$ are the lattice constants along the $x$- and $y$-directions, respectively. For the RSOC $\alpha_{\bf k}$ we assume an isotropic SOC strength, $\alpha_R$ for simplicity.

\begin{figure}
\includegraphics[scale=0.07]{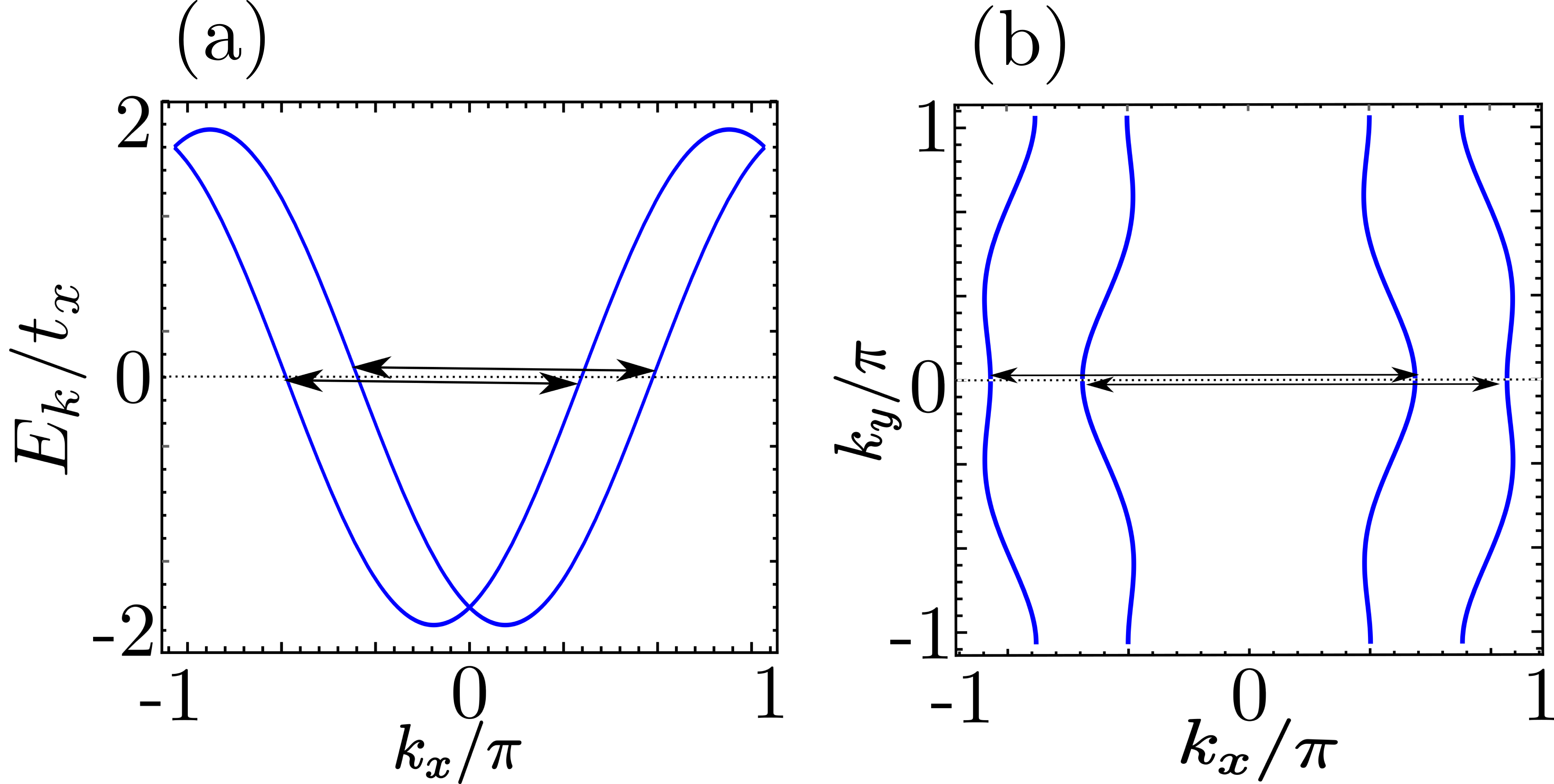}
\caption{(Color online) FS topology. ({a}) Non-interacting RSOC split bands are plotted along $k_x$ with $k_y=0$. Black horizontal arrows show the nesting vectors. ({b}) We show the nesting on the quasi-1D FS.}
\label{fig2}
\end{figure}

The non-interacting dispersion with RSOC is shown in Fig.~\ref{fig2}(a), with two horizontal arrows dictating the ${\bf Q}$ nesting vectors connecting the two helical bands. For our numerical calculations, we use $t_y/t_x=0.2$, $\xi_{\rm F}=0$, and $\alpha_R=-1.25/t_x$, which are realistic parameters for Bi-surface state grown on Ag thin films.\cite{SODW} For Bi- and Pb- atomic wires with one monolayer coverage, the intrinsic value of the FS nesting is $\sim(0.4 2\pi/a,0)$\cite{SODW,PbFS}. Starting from this band parameter, we estimate that the required chemical potential shift to obtain the $(\pi,0)$ nesting is about 1.74 $t_x $, which can be achieved with chemical doping or gating or varying thickness, among others.

The interaction term responsible for the emergence of the QSHDW can be sought from onsite Hubbard, or Hund's coupling or Heisenberg interaction, as shown explicitly in the supplementary material (SM).\cite{SM} Here we use a generalized form as
\begin{eqnarray}
H_{\rm int} = g\sum_{\substack{{\bf k}_1-{\bf k}_4,\\ \sigma_1-\sigma_4}}c^{\dag}_{{\bf k}_1,\sigma_1}c_{{\bf k}_2,\sigma_2}c^{\dag}_{{\bf k}_3,\sigma_3}c_{{\bf k}_4,\sigma_4},
\end{eqnarray}  
where $g$ is the strength of the onsite interaction. $c^{\dag}_{{\bf k},\sigma}$ ($c_{{\bf k},\sigma}$) is the creation (annihilation) operator for an electron with Bloch momentum ${\bf k}$, and spin $\sigma=\pm$.

We define a four component Nambu-Gor'kov spinor $\Psi_{{\bf k}}=(c_{{\bf k},\uparrow}, c_{{\bf k},\downarrow}, c_{{\bf k}+{\bf Q},\uparrow}, c_{{\bf k}+{\bf Q},\downarrow})$. For the particular type of nesting depicted in Figs.~\ref{fig2}(a-b), one singlet and two possible triplet order parameters which can develop as:

\begin{eqnarray}
\text{Singlet:}\left\langle O_1 \right\rangle &=& \sum_{\bf k} \left\langle \bar{\Psi}_{\bf k}\left|\Gamma_1 d_{1{\bf k}}\right|{\Psi}_{\bf k}\right\rangle,\\
\text{Triplet:}\left\langle O_2 \right\rangle &=& \sum_{\bf k} \left\langle \bar{\Psi}_{\bf k}\left|\Gamma_2 d_{2{\bf k}}+ \Gamma_3 d_{3{\bf k}}\right|{\Psi}_{\bf k}\right\rangle,\\
\left\langle O_3 \right\rangle &=& \sum_{\bf k} \left\langle \bar{\Psi}_{\bf k}\left|\Gamma_4 d_{4{\bf k}}\right|{\Psi}_{\bf k}\right\rangle,
\label{OP}
\end{eqnarray}
where the Dirac $\Gamma$-matrices have the representation $\Gamma_{\rm (1,2,3,4,5,6,7)}$ =($\tau_y\otimes\sigma_y$,$\tau_x\otimes\sigma_x$, $\tau_x\otimes\sigma_y$, $\tau_x\otimes\sigma_z$, $\tau_z\otimes \mathbb{I}$, $\mathbb{I}\otimes\sigma_x$, $\tau_z\otimes\sigma_y$) in the same spinor $\Psi$. $\tau_i$, and $\sigma_i$ are the $2\times 2$ Pauli matrices in the sublattice and spin basis, respectively, and $\mathbb{I}$ is the 2$\times$2 identity matrix. Except $\Gamma_{1}$ and $\Gamma_{5}$, all other $\Gamma$ matrices here are odd under TR symmetry. Here, we are interested only in the TR invariant order parameters for $Z_2$ topological consequence. Therefore, the TR invariance of these order parameters requires that the structure factor $d_{i{\bf k}}$ must complement the symmetry of the corresponding $\Gamma_i$ matrices under TR symmetry. Therefore $d_{1{\bf k}}$ for singlet state must be even under TR symmetry, while all three $d_{2,3,4}$ for the triplet states must be odd under TR symmetry. In what follows, the order parameters can be either even parity and spin singlet or odd parity and spin triplet. This is also consistent with the  fermionic antisymmetric property of the order parameters. 

These order parameters introduce  electronic gap terms as $\Delta_i = g\left\langle O_i \right\rangle$. All order parameters govern nontrivial topological phase as to be shown later. For the singlet case, we take $\Delta_{1{\bf k}}=\Delta_{10}$ ($s$-wave) without loosing generality. For the triplet gaps $\Delta_{2,3}$, we find through self-consistent solution (see supplementary materials\cite{SM}) that $\Delta_2$ has higher prosperity to form and possesses larger amplitude than the $\Delta_3$ term. Henceforth,  we thus consider only the $\Delta_2$ term for the triplet case.  We consider a $p$-wave form factor for the odd parity term as $\Delta_{2{\bf k}}=\Delta_{20}\sin{(k_x a)}$.  We note that the essential topological character deduced here does not depend on the form factor, which will be clearer below. At ${\bf Q}=(\pi,0)\text{ or }(0,\pi)$, the mean-field Hamiltonian can be fully expressed in terms of the Dirac matrices as (for singlet):
\begin{eqnarray}
H_1(\textbf{k})=\xi_\textbf{k}^+{\rm I}_{4\times 4} +\xi_\textbf{k}^{-}\Gamma_5 +\alpha^{\prime}_\textbf{k}\Gamma_6+\alpha^{\prime\prime}_\textbf{k}\Gamma_7 +\Delta_{10} \Gamma_1,\\
\text{and eigenvalues:}~E_{1k}=\xi_k^+\pm\sqrt{(\xi_k^{-}\pm|\alpha_k|)^2+\Delta^2_{10}},
\label{Ham}
\end{eqnarray}
and for triplet:
\begin{eqnarray}
H_2(\textbf{k})=\xi_\textbf{k}^+{\rm I}_{4\times 4} +\xi_\textbf{k}^{-}\Gamma_5 +\alpha^{\prime}_\textbf{k}\Gamma_6+\alpha^{\prime\prime}_\textbf{k}\Gamma_7 +\Delta_{2{\bf k}} \Gamma_2,\\
\text{and eigenvalues:} E_{2k}=\xi_k^+\pm|\alpha_k|\pm\sqrt{(\xi_k^{-})^2+\Delta^2_{2,{\bf k}}}.
\label{Ham2}
\end{eqnarray}
Here $\xi_{\bf k}^{\pm}=(\xi_{\bf k}\pm\xi_{{\bf k}+{\bf Q}})/2$, and $\alpha_{\bf k}^{\prime}$, and $\alpha_{\bf k}^{\prime\prime}$ are the real and imaginary parts of the RSOC ($\alpha_k$).  
In analogy with the Dirac Hamiltonian, we can easily recognize that $\xi^{-}_{\bf k}$ gives the Dirac mass term which controls the topological phase transition, while $\Delta_{\bf k}$ helps open  electronic gap between the opposite chiral states.

Few remarks are in order about why the presnt mean-field model gives correct result in such quasi-1D systems. In quasi-1D systems, one may expect that a Luttinger liquid theory might be more appropriate. However, experimentally it is demonstrated that at finite temperature and in the presence of impurity scattering, the quantitative difference between the Luttinger liquid and Fermi liquid behavior is small and often undetectable.\cite{Wang} Therefore, a Fermi liquid like physics with mean-field order parameter can be used here. Moreover, in the weak coupling region, quantum fluctuations are Fermi liquid like, i.e. it scales quadratically with energy. Such weak fluctuations only become appreciable near the quantum critical regime where the gap becomes small. Away from the critical region, the QSHDW order is robust against quantum fluctuations.

{\it  Electronic} insulator. For a pure 1D case ($t_y/t_x\rightarrow 0$), any infinitesimally small value of $\Delta$ produces an insulating band gap. As the FS warping increases with increasing $t_y/t_x$, some parts of the FS (which are not nested by ${\bf Q}$) remain ungapped for small values of $\Delta$ (topological invariant may still be defined for the cases with small FS pockets, giving rise to QSHDW semimetals). With larger $\Delta$, insulating gap appears. The critical value of $\Delta$ required for the insulating state increases with increasing $t_y/t_x$. 

\begin{figure}[t]
	\includegraphics[scale=0.11]{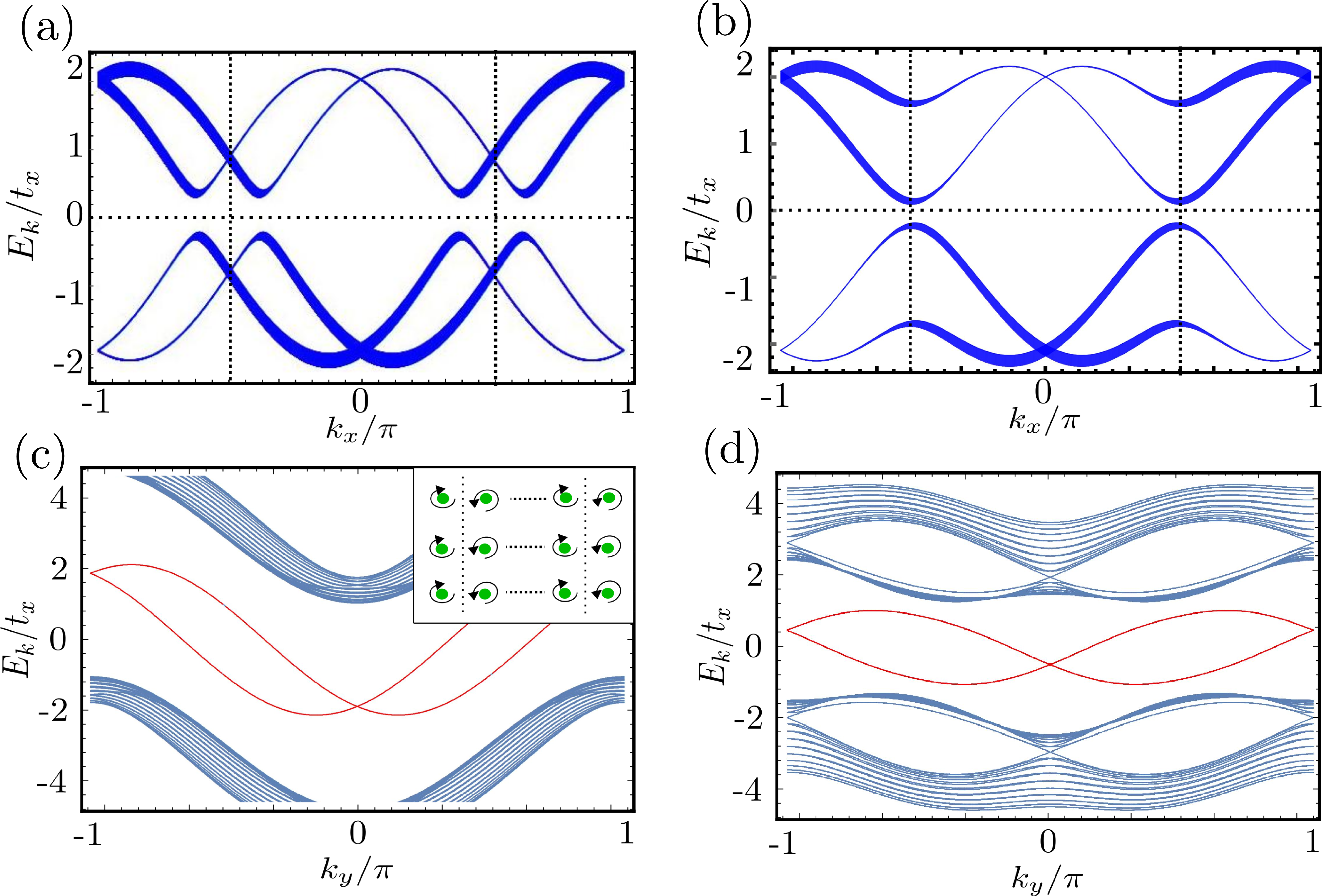}
	\caption{(Color online) Electronic dispersion and Edge states in quasi-1D strip geometries.(a-b) We plot the  Electronic band structure at $k_y$ = 0 for singlet and triplet states, respectively. The width of each line dictates the corresponding  Electronic weight in the QSHDW state. The vertical dashed lines give the RBZ boundaries.(c-d) Edge states in quasi-1D strip geometries for the singlet and triplet state, respectively. We show the spectrum of the interacting quasi-1D QSHDW in a strip geometry (inset).}
\label{fig3}
\end{figure}

In Figs.~\ref{fig3}(a-b), we demonstrate the electronic dispersion for a QSHDW triplet (singlet)insulator. The vertical width of each line in Fig.~\ref{fig3}(a-b) dictates the  electronic weight associated with the main bands (thickness of the line corresponds to the contribution from first reduced BZ (RBZ)). As the main and shadow bands possess different spin-orbit chirality (due to $\alpha_{k+Q}=\alpha_{k}^*$), the emergence of QSHDW order is naturally accompanied by chirality inversion at the TRS momenta. In the present QSHDW theory, due to non-collinearity of the spin coming from the SOC, the spin expectation value of two different bands at each sublattice cancels each other, and thus the system presenves TR symmetry.

{\it Topological properties.} For the calculation of  topological invariants in single particle picture (also applicable to mean field electronic bands), Kane and Mele proposed the concept of `TR polarization'. This is a Z$_2$ analog of the charge polarization for integer quantum hall state\cite{KaneMele}. TR polarization depends on the number of times an electron exchanges its `TR partner' between its Bloch state, $\psi_n({\bf k})$, and its TR conjugate $\psi^{\dag}_m(-{\bf k})$ in half of the BZ. This is essentially quantified by the Pfaffian of a matrix with components $w_{mn}({\bf k})=\langle \psi_m(-{\bf k})|T|\psi_n({\bf k}) \rangle$, where $T=i\mathbb{I}\otimes\sigma_y K$ ($K$ is the complex conjugate operator) is the TR operator, and $n$ and $m$ are valence band indices. Relative sign of Pf[$w(k)$] between any two TR invariant k-points becomes opposite if the electron switches its TR partner odd number of times in traversing between them. This in other words implies that the Pf[$w(k)$] vanishes at odd number of momenta in between the two high-symmetric k-points.\cite{KaneMele} The $Z_2$ invariant $\nu$ is defined as:
\begin{equation}
\nu=\frac{1}{2\pi i}\int_L d{\textbf{k}}.\nabla_{\textbf{k}}\log\left[P\left({\textbf{k}}\right)+i\delta\right]
\label{TRP}
\end{equation}
where L covers half the BZ. As $P({\bf k}^*_i)=0$, the residue theory dictates that $\nu=1$. If there are odd number of $P(k)=0$, one obtains $\nu=1$ (modulo 2), otherwise, $\nu=0$. According to Kane-Mele criterion, there are three $Z_2$ invariants in 2D: ($\nu_0$:$\nu_1\nu_2$), where $\nu_0$ is the net $Z_2$ invariant giving a strong topological insulator, while $\nu_{1,2}$ are the weak topological invariants representing odd number of band inversions along the $x$ and $y$ directions, respectively.  

In the present 1D case, the chirality or the TR polarizibility is reversed along the direction of the nesting. For both singlet and triplet cases, we find that Pf[$w$] changes sign when going from $k_x=0$ to $k_x=\pi$, and it vanishes at $k_x=\pi/2$, but not in the perpendicular directions. Therefore, the system possess a strong $Z_2$ topological invariant ($\nu_1=1$) along this direction (in 1D), but a weak topological insulator in 2D with invariants (0:10). This behavior also makes our model distinct from the Kane-Mele model of the QSH insulator in graphene which is defined by $Z_2$ invariant (1:00).

{\it Boundary state.} Due to the bulk boundary correspondence, non-trivial $Z_2$ invariant implies the existence of zero energy edge states as long as the TR symmetry is held. The present system resembles a Su-Schrieffer-Heeger ~\cite{SSH} type model in 1D if we map the two atoms with opposite chirality  in larger unit cell as two sublattices. Therefore, the topological invariant in the bulk dictates a single end state inside the gap. The end state is localized at the two ends of the lattice in the nesting direction (here $x$-direction), but disperses along the $y$-direction. They are further split by the RSOC.    

To show the behavior of these edge states, we investigate a strip geometry, see {\it inset} to Fig.~3(d) , with open boundary condition along the $x$-direction while keeping the periodic boundary condition along the $y$-direction. Splitting the corresponding (triplet) Hamiltonian into three parts as $H_{\rm strip}=H_1+H_2+H_{12}$, where $H_1$ and $H_2$ are the non-interacting terms in the first and second RBZ, while $H_{12}$ is the interaction term, we get 

\begin{eqnarray}
\label{edge1}
H_1&=&\sum_{k_y,j,\sigma}^{'}\left[-2t_y\cos{(k_y)}c_{k_y,j,\sigma}^\dagger c_{k_y,j,\sigma}-t_x c_{k_y,j,\sigma}^\dagger c_{k_y,j\pm 1,\sigma}\right.\nonumber\\
&&\left.+\alpha_R \sin{(k_y)}c_{k_y,j,\sigma}^\dagger c_{k_y,j,\bar{\sigma}}-\lambda\frac{\alpha_R}{2} c_{k_y,j,\sigma}^\dagger c_{k_y,j+\lambda,\bar{\sigma}}\right],\\
H^s_{12}&=&\Delta_{10}\sum_{k_y,j.\sigma}^{'}\left[e^{i Q_x j}c_{k_y,j,\sigma}^\dagger c_{k_y,j,\bar{\sigma}}\right.\nonumber\\
&&\left.+e^{-i Q_x j}c_{k_y,j,\sigma}^\dagger c_{k_y,j,\bar{\sigma}}\right],\\
H^{t}_{12}&=&-i\Delta_{20}/2\sum_{k_y,j,\sigma}^{'}\left[ e^{-iQ_x(j+1)}c_{k_y,j,\sigma}^\dagger c_{k_y,j+1,\bar{\sigma}}\right.\nonumber\\
&&\left.- e^{-iQ_x(j-1)}c_{k_y,j,\sigma}^\dagger c_{k_y,j-1,\bar{\sigma}}+\textrm{ h.c.}\right].
\label{edge2}
\end{eqnarray}

Here $H_2 = H_1({\bf k}\rightarrow {\bf k}+{\bf Q})$. The index $\lambda=\pm 1$ takes care of the fact that for the RSOC, the nearest neighbor (spin-flip) hopping along $\pm{\bf r}$ directions have opposite sign. $j$ is the lattice site index along the $x$-direction, and prime over summation indicates that it is restricted within the corresponding RBZ. $H^{t/s}_{12}$ corresponds to the  triplet/singlet case. Also, 1$^{st}$ $c$ in H$_{12}^{t/s}$ belongs to $k$ sublattice while 2$^{nd}$ $c$ belongs to $(k+Q)$ sublattice. The eigenvalues of $H_{\rm strip}$ are plotted in Fig.~\ref{fig3} with $\Delta_{0}=1.48 t_x (3.3 t_x)$ for triplet(singlet). This gap value requires an interaction strength of $g\approx 3.3 t_x (5.0 t_x)$. It should be noted that the interaction strength chosen to show the edge state is much higher than the value required to open the insulating gap. For each 1D strip, $\nu_1=1$ invariant dictates zero energy end states (Zak phase). The nearest neighbor end states are coupled to each other by RSOC, and thus are split at all $k_y$ values except at the TR invariant points. Since the bulk system is a weak topological insulator, the boundary states are not immune to perturbations, as also evident from the presence of even number of Dirac nodes in the BZ. 

{\it 2D extension.} Finally, we explore a 2D system in which we explicitly include both nestings $Q_x=(\pi,0)$, and $Q_y=(0,\pi)$, which makes the Hamiltonian in Eq.~[5 \& 7] a $6\times6$ one. In such case, the topological properties become difficult to deduce analytically. Numerically, we find that Pf[$w$] changes sign every time while going from one TRIM point to another, in both $x$, and $y$-directions, giving rise to the weak $Z_2$ invariant (0:11), a 2D QSHDW insulator.

{\it Conclusions.} We presented the theory of a new state of matter, called QSHDW state, which is a spontaneous symmetry breaking quantum phase associated with a non trivial $Z_2 $ invariant. Designing and synthesis of quasi-2D atomic quantum wires have become a routine laboratory exercise, and it has been extensively shown that  both intrinsic and extrinsic tunings of electronic properties, SOC and Coulomb interaction are very easy in such geometry.\cite{PbNC} In fact, the FS nesting between different helical states is observed in a number of quasi-1D,\cite{PbFS} and 2D systems.\cite{Bi2D} Moreover, it is shown that the FS nesting properties, RSOC as well as the charge screening process can be monitored by varying sample thickness and substrate.\cite{PbFS,Bi2D}  In this connection, ferroelectric or polar substrates can also have versatile role to enhance SOC and interaction strength. 

1D SOC is recently observed in optical lattice, where our idea can also be explored with the existing setups. From theoretical perspective, the generalization of the proposed topological phase to higher dimensional FS with the same nesting condition along all directions is possible. For example, non-centrosymmetric heavy-fermion materials would be potential candidates to explore large SOC an interaction. Therefore, we envision that the emergence of QSHDW insulator may open a new area in the field of interaction induced TIs.

\begin{acknowledgements}
The work is facilitated by the computer cluster facility at the Department of Physics at the Indian Institute of Science. We acknowledge funding from the DST under young research scientist award given through the SERB.
\end{acknowledgements}

\pagebreak[1]
\widetext

\begin{center}
	\textbf{\Large Supplemental Materials: Quantum spin Hall density waves of correlated fermions}
\end{center}

\setcounter{equation}{0}
\setcounter{figure}{0}
\setcounter{table}{0}
\setcounter{page}{1}
\makeatletter
\renewcommand{\theequation}{S\arabic{equation}}
\renewcommand{\thefigure}{S\arabic{figure}}
\renewcommand{\bibnumfmt}[1]{[S#1]}
\renewcommand{\citenumfont}[1]{S#1}

In this supplementary material, we give details of the derivations corresponding to various terms presented in the main text. In Sec.~I, we show that how the spin orbit density wave order parameter can be derived from the Hubbard interaction, Heisenberg interaction and Hund's coupling. We self-consistently calculate various order parameters for as a function of interaction strength $g$ in Sec.~II. The full forms of the $\Gamma$ matrices representation are given in Sec.~III. Calculation of Z$_2$ invariant from parity operator is shown in Sec.~IV.

\section{$\textup{\uppercase\expandafter{\romannumeral 1}}$. Derivation of the order parameter}
\textbf{$\textup{\uppercase\expandafter{\romannumeral 1}}$A. Hubard Interaction}

\begin{equation}
H_{\rm int}=\frac{U}{N}\sum_i n_{i\uparrow}n_{i\downarrow}=\frac{U}{N}\sum_i c_{i\uparrow}^\dagger c_{i\uparrow}c_{i\downarrow}^\dagger c_{i\downarrow},
\end{equation}
where $i,j$ are the site index and $N$ is the total number of sies. Taking the Fourier transformation to the momentum space, we get
\begin{eqnarray}
H_{\rm int}&=&\frac{U}{N}\sum_i \left(\frac{1}{N^2}\sum_{k,k',k'',k'''}e^{i({\bf k}-{\bf k}'+{\bf k}''-{\bf k}''').{\bf r}_i}c_{k\uparrow}^\dagger c_{k'\uparrow}c_{k''\downarrow}^\dagger c_{k'''\downarrow}\right),\nonumber\\
&=&\left.\frac{U}{N^2}\sum_{k,k',k''}c_{k\uparrow}^\dagger c_{k'\uparrow}c_{k''\downarrow}^\dagger c_{k'''\downarrow}\right|_{k'''=k-k'+k''}
\end{eqnarray}

(1) For $k'=k$, and $k''=k+Q$ we get
\begin{equation}
H_{\rm int}^1=g\sum_k c_{k\uparrow}^\dagger c_{k\uparrow}c_{k+Q\downarrow}^\dagger c_{k+Q\downarrow},
\end{equation}
where $g=U/N^2$. Using the fermionic anti-commutation relations we can rearrange the operators in $H_{int}$ to get
\begin{equation}
H_{\rm int}^1=g\sum_k\left( c_{k\uparrow}^\dagger c_{k\uparrow}-c_{k\uparrow}^\dagger c_{k+Q\downarrow}c_{k+Q\downarrow}^\dagger c_{k\uparrow}\right).
\end{equation} 
The first term gives the Hartee interaction which is neglected here (usually density-functional theory based calculation incorporates this term). Expanding the Hamiltonian in terms of the mean-field order parameter $\Delta_1$, we get
\begin{eqnarray}
%\begin{aligned}
%\begin{split}
H_{\rm int}^{1}&=&-g\sum_k \left( \langle c_{k\uparrow}^\dagger c_{k+Q\downarrow}\rangle c_{k+Q\downarrow}^\dagger c_{k\uparrow}\right.\nonumber\\
&& \left.+ c_{k\uparrow}^\dagger c_{k+Q\downarrow} \langle c_{k+Q\downarrow}^\dagger c_{k\uparrow}\rangle
-\langle c_{k\uparrow}^\dagger c_{k+Q\downarrow}\rangle\langle c_{k+Q\downarrow}^\dagger c_{k\uparrow}\rangle\right).\nonumber\\
%\end{split}
%\end{aligned}
\end{eqnarray}

(2) If we take $k'$and $k''$=$k+Q$ in Eq.~(S2), we get
\begin{equation}
H_{\rm int}^{2}=g\sum_k c_{k\uparrow}^\dagger c_{k+Q\uparrow}c_{k+Q\downarrow}^\dagger c_{k\downarrow},
\end{equation}
Expanding the Hamiltonian in terms of the mean-field order parameter $\Delta_2$, we get
\begin{equation}
\begin{aligned}
\begin{split}
H_{\rm int}^{2}=g\sum_k&\left( \langle c_{k\uparrow}^\dagger c_{k+Q\uparrow}\rangle c_{k+Q\downarrow}^\dagger c_{k\downarrow}+ c_{k\uparrow}^\dagger c_{k+Q\uparrow} \langle c_{k+Q\downarrow}^\dagger c_{k\downarrow}\rangle-\right.\\
& \left.\langle c_{k\uparrow}^\dagger c_{k+Q\uparrow}\rangle\langle c_{k+Q\downarrow}^\dagger c_{k\downarrow}\rangle\right).
\end{split}
\end{aligned}
\end{equation}

\vspace{1cm}\textit{}
\textbf{$\textup{\uppercase\expandafter{\romannumeral 1}}$B. Heisenberg Interaction}

\begin{equation}
H_{\rm int}=JS_i.S_j,
\end{equation}
where $i,j$ are site index and

\begin{equation}
\begin{aligned}
\begin{split}
{\bf S}_{i}&=c_i^\dagger {\bf \sigma}c_i,\\
S_{ix}&=c_{i\uparrow}^\dagger c_{i\downarrow}+c_{i\downarrow}^\dagger c_{i\uparrow},\\
S_{iy}&=-i\left(c_{i\uparrow}^\dagger c_{i\downarrow}-c_{i\downarrow}^\dagger c_{i\uparrow}\right),\\
S_{iz}&=c_{i\uparrow}^\dagger c_{i\uparrow}-c_{i\downarrow}^\dagger c_{i\downarrow},
\end{split}
\end{aligned}
\end{equation}
This gives
\begin{equation}
H_{\rm int}=J(2c_{i\uparrow}^\dagger c_{i\downarrow}c_{j\downarrow}^\dagger c_{j\uparrow}+2c_{i\downarrow}^\dagger c_{i\uparrow}c_{j\uparrow}^\dagger c_{j\downarrow}-c_{i\uparrow}^\dagger c_{i\uparrow}c_{j\downarrow}^\dagger c_{j\downarrow}-c_{i\downarrow}^\dagger c_{i\downarrow}c_{j\uparrow}^\dagger c_{j\uparrow}),
\end{equation}
as seen in the Hubbard term, the above terms will also lead to the similar terms on taking the fourier transformation and hence leads to the similar type of order parameter.

\vspace{1cm}
\textbf{$\textup{\uppercase\expandafter{\romannumeral 1}}$C. Hund's Coupling}

\begin{equation}
H_{\rm int}=J_H S_\alpha .S_\beta,
\end{equation}
where $\alpha,\beta$ are band indices. Since Hund's coupling is very much similar to Heisenberg interaction hence this will also lead to the similar type of order parameter.

\vspace{1cm}
\textbf{$\textup{\uppercase\expandafter{\romannumeral 1}}$D. General Interaction Hamiltonian}

\begin{eqnarray}
H_{\rm int} = g\sum_{\substack{{\bf k}_1-{\bf k}_4,\\ \sigma_1-\sigma_4}}c^{\dag}_{{\bf k}_1,\sigma_1}c_{{\bf k}_2,\sigma_2}c^{\dag}_{{\bf k}_3,\sigma_3}c_{{\bf k}_4,\sigma_4},
\end{eqnarray}  
where $g$ is $U$ for Hubbard interaction, $J$ for Heisenberg interaction and $J_H$ for Hund's coupling. 
%The momentum and spin conservations imply that ${\bf k}_1+{\bf k}_3={\bf k}_2+{\bf k}_4$, and $\sigma_1+\sigma_3=\sigma_2+\sigma_4$, with $\sigma=\pm$. For Hubbard interaction, we obtain $\sigma_1=\sigma_2=-\sigma_3=-\sigma_4$, while for Hund's coupling and Heisenberg interaction with in-plane spin, we get $\sigma_1=-\sigma_2=\sigma_3=-\sigma_4$. For the non-interacting RSOC bands Fig. 2(a), the FS nesting ${\bf Q}$ between the two helical states further constraints the momentum, and spin indices to be ${\bf k}_1={\bf k}_2={\bf k}$, ${\bf k}_3={\bf k}_4={\bf k}+{\bf Q}$, and $\sigma_3=-\sigma_2=\sigma$, and  $\sigma_1=-\sigma_4$, and the choice between $\sigma_1$ and $\sigma_2$ is subjected to the interaction term and does not impact the result. 

\section{$\textup{\uppercase\expandafter{\romannumeral 2}}$. Self-consistent gap equations}

In this section we will show how the two gap terms ($\Delta_{1,2}$) changes with the interaction strength by solving it self-consistently. The mean-field Hamiltonians corresponding to the two order parameters are
$$H_1=\begin{bmatrix}
\xi_k & \alpha_k & 0 & \Delta_1\\
\alpha_k^* & \xi_k & \Delta_1 & 0\\
0 & \Delta_1 & \xi_{k+Q} & \alpha_k^*\\
\Delta_1 & 0 & \alpha_k & \xi_{k+Q}
\end{bmatrix},$$
$$H_2=\begin{bmatrix}
\xi_k & \alpha_k & \Delta_2 & 0\\
\alpha_k^* & \xi_k & 0& -\Delta_2 \\
\Delta_2 &0 &  \xi_{k+Q} & \alpha_k^*\\
0 &-\Delta_2 &  \alpha_k & \xi_{k+Q}
\end{bmatrix}.$$
So we can evaluate $\Delta_{1(2)}$ self-consistently by finding out the expectation value of $\Gamma_1$ ($\Gamma_3$), i.e, $\sum_n\bra{n}\Gamma_1\ket{n}$ $\left(\sum_n\bra{n}\Gamma_3\ket{n}\right)$ where $\ket{n}$ are the eigenstates of H below Fermi level. The corresponding result is shown in Fig.~S1.

\begin{figure}[ht]
	\includegraphics[scale=0.5]{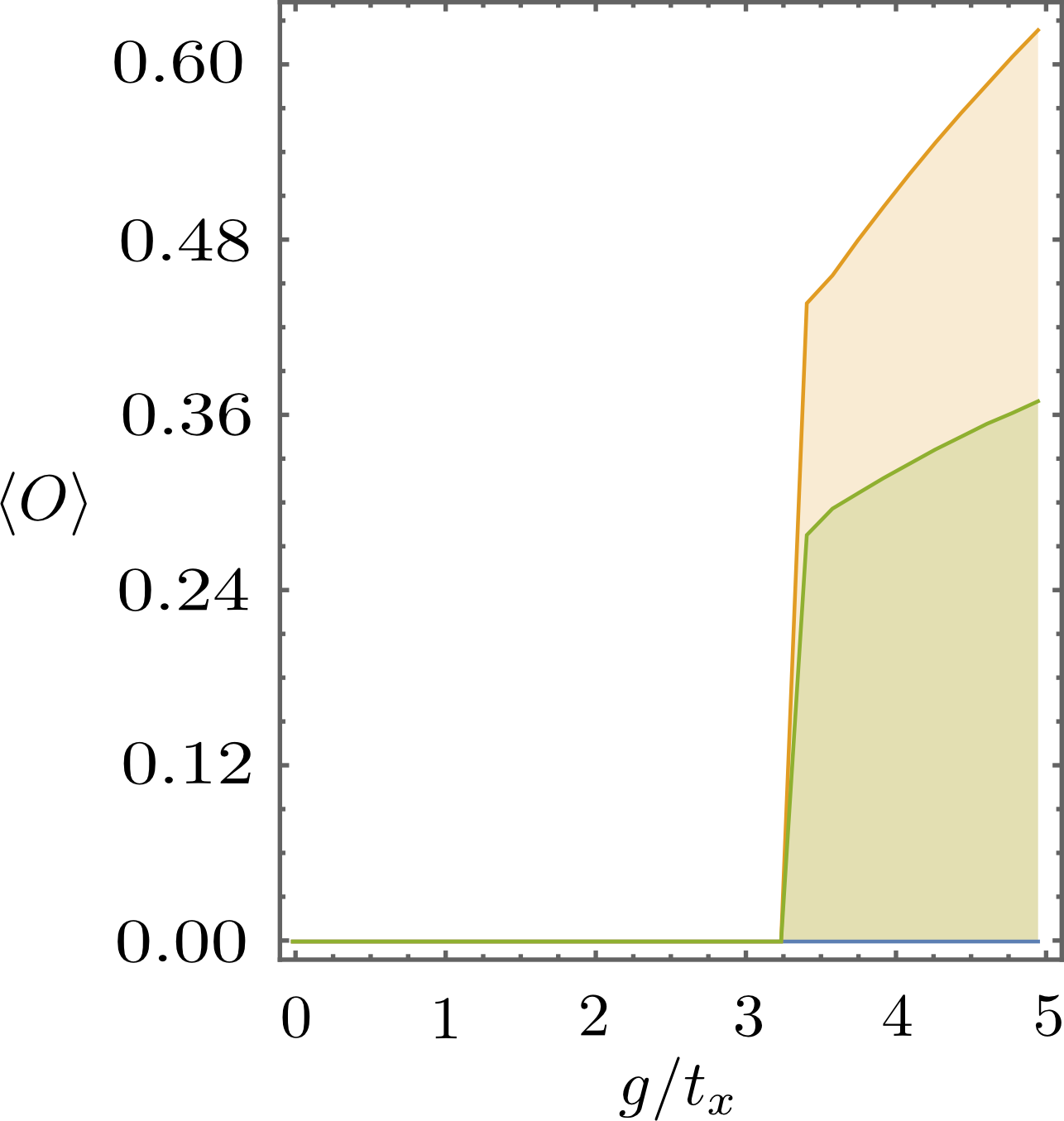}
	\caption{\label{fig:epsart}We show the variation of order parameter $\Delta_1$ and $\Delta_2$ (defined in Eq.~1 in the main text) as a function of interaction strength $g$. Orange curve shows the order parameter $\Delta_1$ while blue curve shows that for $\Delta_2$.}
\end{figure}

\section{$\textup{\uppercase\expandafter{\romannumeral 3}}$. Full Gamma matrices}

$$\Gamma_0=\mathbb{I}\otimes\mathbb{I}=\begin{bmatrix}
1 & 0 & 0 & 0\\
0 & 1 & 0 & 0\\
0 & 0 & 1 & 0\\
0 & 0 & 0 & 1
\end{bmatrix}$$

$$\Gamma_1=\tau_x\otimes\sigma_x=\begin{bmatrix}
0 & 0 & 0 & 1\\
0 & 0 & 1 & 0\\
0 & 1 & 0 & 0\\
1 & 0 & 0 & 0
\end{bmatrix}$$

$$\Gamma_2=\tau_x\otimes\sigma_y=\begin{bmatrix}
0 & 0 & 0 & -i\\
0 & 0 & i & 0\\
0 & -i & 0 & 0\\
i & 0 & 0 & 0
\end{bmatrix}$$

$$\Gamma_3=\tau_x\otimes\sigma_z=\begin{bmatrix}
0 & 0 & 1 & 0\\
0 & 0 & 0 & -1\\
1 & 0 & 0 & 0\\
0 & -1 & 0 & 0
\end{bmatrix}$$

$$\Gamma_4=\tau_z\otimes\mathbb{I}=\begin{bmatrix}
1 & 0 & 0 & 0\\
0 & 1 & 0 & 0\\
0 & 0 & -1 & 0\\
0 & 0 & 0 & -1
\end{bmatrix}$$

$$\Gamma_5=\mathbb{I}\otimes\sigma_x=\begin{bmatrix}
0 & 1 & 0 & 0\\
1 & 0 & 0 & 0\\
0 & 0 & 0 & 1\\
1 & 0 & 1 & 0
\end{bmatrix}$$

$$\Gamma_6=\tau_z\otimes\sigma_y=\begin{bmatrix}
0 & -i & 0 & 0\\
i & 0 & 0 & 0\\
0 & 0 & 0 & i\\
0 & 0 & -i & 0
\end{bmatrix}$$


\begin{thebibliography}{100}
\bibitem{TIreviewTD}A. Bansil, H. Lin, and T. Das, Rev. Mod. Phys. {\bf{88}}, 021004 (2016).
\bibitem{TIreviewSCZ} X.-L. Qi and S.-C. Zhang, Rev. Mod. Phys. {\bf{83}}, 1057 (2011).
\bibitem{TIreviewCK} M. Z. Hasan and C. L. Kane, Rev. Mod. Phys. {\bf{82}}, 3045 (2010).
\bibitem{TMI} S. Raghu, X.-L. Qi, C. Honerkamp, and S.-C. Zhang, Phys. Rev. Lett. {\bf{100}}, 156401 (2008).
\bibitem{TAI} R. S. K. Mong, A. M. Essin, and J. E. Moore, Phys. Rev. B {\bf{81}}, 245209 (2010).
\bibitem{TKI} M. Dzero, K. Sun, V. Galitski, and P. Coleman, Phys. Rev. Lett. {\bf{104}}, 106408 (2010).
\bibitem{TDI} J. Li, R.-L. Chu, J. K. Jain, and S.-Q. Shen, Phys. Rev. Lett. {\bf{102}}, 136806 (2009).
\bibitem{TAFM} R. S. K. Mong, A. M. Essin, and J. E. Moore, Phys. Rev. B. {\bf{81}}, 245209 (2010).
\bibitem{HgTeTh} B. A. Bernevig, T. L. Hughes, and S.-C. Zhang, Science {\bf{314}}, 1757 (2006).
\bibitem{HgTeExp} M. König, S. Wiedmann, C. Brüne, A. Roth, H. Buhmann, L. W. Molenkamp, X.-L. Qi, and S.-C. Zhang,Science {\bf{318}}, 766 (2007).
\bibitem{InAsGaSb} I. Knez, R.-R. Du, and G. Sullivan, Phys. Rev. Lett. {\bf{107}}, 136603 (2011).
\bibitem{BiSbTh} J. C. Y. Teo, L. Fu, and C. L. Kane, Phys. Rev. B {\bf{78}}, 045426 (2008).
\bibitem{BiSbExp} D. Hsieh, D. Qian, L. Wray, Y. Xia, Y. S. Hor, R. Cava,and M. Z. Hasan, Nature {\bf{452}}, 970 (2008).5
\bibitem{BiSeTh} H. Zhang, C.-X. Liu, X.-L. Qi, X. Dai, Z. Fang, and S.-C. Zhang, Nat. Phys. {\bf{5}}, 438 (2009).
\bibitem{BiSeExp} Y. Xia, D. Qian, D. Hsieh, L. Wray, A. Pal, H. Lin, A. Bansil, D. Grauer, Y. Hor, R. Cava, et al., Nat. Phys. {\bf{5}}, 398 (2009)
\bibitem{BiTeExp} Y. Chen, J. Analytis, J.-H. Chu, Z. Liu, S.-K. Mo, X.-L.Qi, H. Zhang, D. Lu, X. Dai, Z. Fang, et al., Science {\bf{325}},178 (2009).
\bibitem{SmB6}J. Jiang, S. Li, T. Zhang, Z. Sun, F. Chen, Z. Ye, M. Xu, Q. Ge, S. Tan, X. Niu, et al., Nat. commun. {\bf{4}} (2013).
\bibitem{YbB6}T.-R. Chang, T. Das, P.-J. Chen, M. Neupane, S.-Y. Xu,M. Z. Hasan, H. Lin, H.-T. Jeng, and A. Bansil, Phys. Rev. B {\bf{91}}, 155151 (2015).
\bibitem{PbFS} C. Tegenkamp, D. Lükermann, H. Pfnür, B. Slomski, G. Landolt, and J. Dil, Phys. Rev. Lett. {\bf{109}}, 266401 (2012).
\bibitem{PbNC} C. Brand, H. Pfnür, G. Landolt, S. Muff, J. Dil, T. Das, and C. Tegenkamp, Nat. Commun. {\bf{6}} (2015).
\bibitem{SODW}T. Das, Phys. Rev. Lett. {\bf{109}}, 246406 (2012).
\bibitem{SM} See supplementary materials for the details of the calculations.
\bibitem{Wang} D.-W. Wang, A. Millis, and S. D. Sarma, Phys. Rev. Lett. {\bf{85}}, 4570 (2000).
\bibitem{KaneMele} C. L. Kane and E. J. Mele, Phys. Rev. Lett. {\bf{95}}, 146802 (2005).
\bibitem{FuKane} L. Fu and C. L. Kane, Phys. Rev. B {\bf{76}}, 045302 (2007).
\bibitem{Bi2D} H. Bentmann, S. Abdelouahed, M. Mulazzi, J. Henk, and F. Reinert, Phys. Rev. Lett. {\bf{108}}, 196801 (2012).
\bibitem{SSH}W. P. Su, J. R. Schrieffer, and A. J. Heeger, Phys. Rev. Lett. {\bf{42}}, 1698 (1979).


\end{thebibliography}

\begin{thebibliography}{100}
	\bibitem{FuKane} L. Fu and C. L. Kane, Phys. Rev. B {\bf{76}}, 045302 (2007).
\end{thebibliography}
\end{document}